# Evidence of long-range ferromagnetic order and spin frustration effects in the double perovskite La$_2$CoMnO$_6$

R. P. Madhogaria, R. Das, E. M. Clements, V. Kalappattil, M. H. Phan*, H. Srikanth*

Department of Physics, University of South Florida, Tampa, Florida 33620, USA

N. T. Dang

Institute of Research and Development, Duy Tan University, Da Nang 550000, Viet Nam

D. P. Kozlenko

Frank Laboratory of Neutron Physics, Joint Institute for Nuclear Research, Dubna 141980, Russia

N. S. Bingham

Department of Applied Physics, Yale University, New Haven, Connecticut 06511, USA

## ABSTRACT

We present a comprehensive study on the magnetic structure, dynamics, and phase evolution in the single-phase double perovskite La$_2$CoMnO$_6$. The mixed valence state due to oxygen deficiency is verified by X-ray photoelectron spectroscopy, and confirms a double ferromagnetic transition observed in DC magnetization. Neutron diffraction reveals that the magnetic structure is dominated by long-range ferromagnetic ordering, which is further corroborated by a critical exponents analysis of the paramagnetic to ferromagnetic phase transition. An analysis of the magnetization dynamics by means of linear and nonlinear ac magnetic susceptibilities marks the presence of two distinct cluster glass-like states that emerge at low temperatures. The isothermal entropy change as a function of temperature and magnetic field ($H$) is exploited to investigate the mechanism of stabilization of the magnetic phases across the $H$-$T$ phase diagram. In the regime of the phase diagram where thermal energy is sufficiently low, regions of competing interactions due to local disorder become stabilized and display glass-like dynamics. The freezing mechanism of clusters is illustrated using a unique probe of transverse susceptibility that isolates the effects of the local anisotropy of the spin clusters. The results are summarized in a new $H\text{-}T$ phase diagram of $La_2CoMnO_6$ revealed for the first time from these data.

KEYWORDS: Double perovskites; Collective magnetism; Magnetic glassy; Magnetic anisotropy

*Corresponding authors: phanm@usf.edu (MHP) and sharihar@usf.edu (HS)

## I. INTRODUCTION

Double perovskite oxides with a general formula $A_2BB'O_6$ (where A is a rare earth or alkaline earth metal, and B and B′ are d-block transition metals) display a wide range of interesting physical properties with composition variations, as two B-site cations allow for novel combinations of different elements [1]. The properties of these oxides are governed by B-site cations, which provide a platform to explore new double perovskite materials and study their electrical and magnetic properties [1][2][3]. The electrical properties of these systems vary from insulator-metal to semiconductor-half metal, and their magnetic properties range from ferromagnetic to spin glass [1]. Extensive research has been done to improve magnetodielectric, magnetoresistance, and magnetocapacitance properties, and to adapt the materials chemistry for large scale production for technological applications [4][5][6]. As double perovskite oxides are promising candidates for novel applications, such as high density recording media, spintronics, magnetic field sensors and infrared detectors [7][8][9][10], it is essential to advance the synthesis and characterization techniques of complex compositions and structures.

One member of the double perovskite oxides, $La_2CoMnO_6$ (LCMO), has gained particular attention because of the presence of the magnetodielectric effect and multiferroic properties [11][12]. LCMO crystallizes in several phases, namely monoclinic ($P2_1/n$ or pseudotetragonal), orthorhombic ($Pbnm$) and rhombohedral ($R3c$) [12], depending on the synthesis route. A paramagnetic (PM) to ferromagnetic (FM) phase transition close to room temperature with insulating behavior is seen in the monoclinic $P2_1/n$ (or pseudotetragonal) phase of this material [6]. The FM insulating properties are governed by the Goodenough-Kanamori rules, with FM coupling arising from a 180º superexchange interaction between two transition metal cations [13]. The magnetic properties of the LCMO system are governed by cation ordering, cation valences and defects, all of which, along with the structural phase composition, depend sensitively on their synthesis conditions [6]. In a study by Dass and Goodenough on the solid state synthesis of LCMO, [6] it was reported that the monoclinic $P2_1/n$ phase was obtained with a saturation magnetization ($M_S$) value of 4.85 $\mu_B$/f.u. when the sample was annealed at 1350 ºC (12 h) in the presence of $O_2$ and cooled at 20 ºC/h. In that same study, the sample annealed at 600 ºC (12 h) in air and cooled at 180 ºC/h alternatively yielded a monoclinic pseudotetragonal with $M_S$ = 3.62 $\mu_B$/f.u. The authors concluded that the high $M_S$ corresponded to a highly atomically ordered sample in which the atomic disorder was attributed to oxygen vacancies rather than antiphase boundaries. Villar et. al. [14] synthesized LCMO using a nitrate decomposition method and obtained a mixed phase of 97% orthorhombic and 3% rhombohedral. They found that the cooling rate had a direct effect on the magnitude of the $M_S$. The sample annealed at 1325 ºC (20 h) and cooled at 20 ºC/h in oxygen had 22.5 % higher $M_S$ than the sample annealed at 1300 ºC (40 h) and cooled at 45 ºC/h in air. In the former sample, the high magnetization ($M_S$ = 5.54 $\mu_B$/f.u.) was related to long-range cationic and valence ordered $Mn^{4+}$/$Co^{2+}$ ions and strong ferromagnetic interactions. In the latter, the decreased value of $M_S$ (4.42 $\mu_B$/f.u.) was attributed to the presence of $Mn^{3+}$/$Co^{3+}$ ions.

The presence of multiple and complicated crystal structures, together with alternating cation valencies and defects, leads to inconsistencies in the magnetic transition temperatures [6][14][15]. The reported PM-FM phase transition temperatures ($T_C$) vary across a broad range from $170 \leq T \leq 235$ K [6][16][17]. These discrepancies are likely due to the presence of oxygen vacancies and formation of antiphase boundaries between specimens. The ordering of $Co^{2+}$ and $Mn^{4+}$ cations, which leads to 180º superexchange interactions, supports a ferromagnetic exchange at higher temperatures, while the oxygen vacancies introduce $Co^{3+}$ and $Mn^{3+}$ ions resulting in atomic disorder and a lowering of $T_C$ [6][13].

Multiple magnetic transitions have been reported in LCMO due to coexisting cationic orders. Guo et al. [15] found that materials synthesized at relatively high and intermediate oxygen pressure had two values of $T_C$ at 230 K and 80 K associated with $Mn^{4+}$-$O^{2-}$-$Co^{2+}$ ferromagnetic superexchange and $Mn^{3+}$-$Co^{3+}$ ferromagnetic vibronic superexchange interactions, respectively. They have also reported the existence of spin glass-like behavior below both $T_C$ values due to the competing character between FM $Mn^{4+}$-O-$Co^{2+}$ and antiferromagnetic (AFM) $Mn^{4+}$-O-$Mn^{4+}$ or $Co^{2+}$-O-$Co^{2+}$ interactions. The spin glass state at zero applied magnetic field ($H$) has also been reported by Wang et al. on the basis of frequency dependent peaks at $T_C$ (220 K) in the real part of the AC susceptibility and slow spin relaxation for $T < T_C$ [18]. Murthy et al. [19] showed that for a monoclinic (pseudotetragonal) phase with two FM transitions at 218 K and 135 K, a frequency dependent peak at 33 K is also present due to the reentry of a frozen state of non-interacting clusters from a state of long-range FM order. However, a direct relationship between oxidation state and cation order/disorder and the definitive answer to the magnetic ground state of LCMO is still a matter of debate. As the exact nature of the magnetic phases, magnetic transitions and phase evolution with respect to field and temperature remain unknown, a thorough understanding may provide a route for designing new materials for novel applications.

In this study, the magnetic properties of a single phase polycrystalline sample of LCMO are characterized by means of DC magnetization, AC susceptibility, magnetic entropy change ($\Delta S_M$), neutron diffraction (ND), and X-ray Photoelectron Spectroscopy (XPS). Our study sheds light on the nature of the magnetic transitions governed by different magnetic interactions and how each phase evolves in this system. Neutron diffraction indicates the existence of long-range FM order for $T < T_C$ down to $T = 20$ K. A systematic analysis of the critical exponents using magnetization data supports the presence of a PM-FM phase transition, in agreement with ND results. AC susceptibility reveals the presence of cluster glass dynamics at low temperature that is further supported by the $\Delta S_M$ measurements, which displays signatures of a magnetic state that competes with the dominant FM phase. The magnetic anisotropy is probed by transverse susceptibility as a function of temperature to investigate the effects of coexisting magnetic states and associated freezing phenomena. The article is organized as follows. Experimental methods are described in Sec. II. Section III. presents results and discussion. In Sec. III A, the structural and cationic ordering of the synthesized sample are confirmed. In Sec. III B, features of the DC magnetization are reviewed. In Sec. III C, the magnetic structure is examined via neutron diffraction, and a critical exponents analysis is performed on the magnetization data. In Sec. III D, short-range ordering due to disorder/defects and its influence on the time dependence and anisotropy is investigated as a major competition to the predominant long-range ordering. Sec. III E presents further analysis and discussion of the phase evolution via the magnetocaloric effect, and the results of the static and dynamic studies are summarized in an $H$-$T$ phase diagram. The conclusion is given in Sec. IV.

## II. EXPERIMENTAL DETAILS

Polycrystalline LCMO was prepared using a conventional sol-gel method. Reagent chemicals lanthanum nitrate hexahydrate ($LaN_3O_96H_2O$), cobalt sulphate heptahydrate ($CoSO_47H_2O$) and manganese nitrate tetrahydrate ($MnN_2O_64H_2O$) were taken as the precursor materials according to their stoichiometric ratios. Citric acid was added as the chelating agent to the stoichiometric mixture of precursors. The entire reaction was carried out in basic medium. The resulting mixture was continuously stirred and heated from $100 - 400$ °C until a viscous gel was formed and decomposed. Final calcination was done at 950 °C for 8 h.

The structural phase analysis was carried out using Bruker AXS powder x-ray diffractometer (XRD) with Cu-K$_\alpha$ radiation. X-ray Photoelectron Spectroscopy (XPS) was performed with a PHI VersaProbe II, in ultra-high vacuum ($\sim 10^{-8}$ torr). Magnetic measurements were performed using a Quantum Design Physical Property Measurement System (PPMS) with the Vibrating Sample Magnetometer (VSM) and AC Measurement System (ACMS) options. Magnetization versus applied magnetic field ($M$ vs. $H$) was measured from $H = 0 - 50$ kOe for temperatures ranging from $T = 10 - 250$ K, and magnetization versus temperature was measured from $T = 10 - 260$ K for $H = 100$ Oe. The first ($\chi_1$), second ($\chi_2$) and third ($\chi_3$) harmonics of AC susceptibility were measured simultaneously for $T = 10 - 250$ K for frequencies $f = 500 - 10000$ Hz with an AC driving field amplitude $h = 5$ Oe. Magnetic entropy change was calculated from magnetization versus field data which was measured up to $H = 30$ kOe and with temperature steps of 5 K from $T = 20 - 200$ K, 1 K from $T = 200 - 232$ K and 2 K from $T = 232 - 244$ K. Transverse susceptibility measurements were performed using a tunnel diode oscillator (TDO) with a resonant frequency of 12 MHz and sensitivity on the order of 10 Hz [20]. This measurement was done for $T = 20 - 120$ K with the applied field from $H = 0 - 50$ kOe. The neutron powder diffraction measurements in the temperature range $T = 20 - 290$ K were performed with the DN-12 diffractometer [21], (IBR-2 pulsed reactor, JINR, Russia). The experimental data were analysed by the Rietveld method using the Fullprof program [22].

## III. RESULTS AND DISCUSSION

### A.    Structural and cationic ordering

The XRD pattern of LCMO is shown in Fig. 1(a). A single monoclinic (P2$_1$/n) phase with no impurities was obtained [6] with crystallite size of 32.5 nm estimated using the Debye Scherrer formula. The inset of Fig. 1(b) shows the SEM image of LCMO particles that are homogenously distributed with an average particle size of 200 nm.    The chemical characterization of the sample was done by Energy-Dispersive-X-ray Spectroscopy (EDS) using a JEOL JSM-6390LV SEM. The EDS spectra show 22.94 wt% of La (SD = 0.66), 10.61 wt% of Co (SD = 0.22), 9.82 wt% of Mn (SD = 0.35) and 56.63 wt% of O (SD = 1.12).

The XPS was recorded at the Mn and Co L$_{2,3}$ edges [Figs. 1(c) and (d), respectively] to confirm the mixed valence states in the transition metal ions. The fit was performed using the CTM4XAS program [23], in which a mixed state of 25% Mn$^{3+}$/Co$^{3+}$ and 75% Mn$^{4+}$/Co$^{2+}$ was calculated. The Mn and Co L$_{2,3}$ spectra [Figs. 1(c) and (d)] were fitted by weighting contributions from Mn$^{4+}$ and Mn$^{3+}$, and Co$^{3+}$ and Co$^{2+}$. The following parameters were used for data collected at the Mn edge: (1) for Mn$^{3+}$, $10D_q = 2$ eV ($\Delta t = 0.05$eV, $\Delta s = 0.4$ eV), $\Delta = 3$ eV, $U_{dd}-U_{pd} = 1$ eV; (2) for Mn$^{4+}$, $10D_q = 2.4$ eV, $\Delta = -3$ eV, $U_{dd}-U_{pd} = 2$ eV. Likewise, the following parameters were used at the Co edge: (1) for Co$^{2+}$, $10D_q = 1$eV, $\Delta = 1$ eV, $U_{dd}-U_{pd} = 1$eV; (2) for Co$^{3+}$, $10D_q = 1.2$ eV, $\Delta = 3$ eV, $U_{dd}-U_{pd} = 1$ eV. For all of the fits, the Slater integrals were reduced to around 80%. In Fig. 1(d), a small peak appears at 804 eV in the experimental data as well as the fit, which is attributed to Co $3d$ – O $2p$ charge transfer [24].

## B.    Static magnetization

The temperature dependence of the zero field cooled (ZFC) and field cooled (FC) magnetization was measured for $T$ = 10 – 270 K at $H$ = 100 Oe, as shown in Fig. 2(a). Divergence is clearly seen between FC and ZFC magnetization curves below 226 K. Apart from the peak at 226 K, a peak at 218 K is also clearly visible in the ZFC magnetization curve. Fig. 2(b), shows $dM/dT$ vs. $T$ for both ZFC and FC magnetization, which further reveals the presence of two anomalies in close proximity to each other. The minimum at 228 K ($T_{F1}$) and the minimum at 220 K ($T_{F2}$) correspond to two different FM interactions. The large deviation between FC and ZFC below 226 K suggests the presence of magnetic frustrations in the system [25]. Fig. 2(c), shows $M$ vs. $T$ measurements at $H$ = 60 kOe. Despite the large applied magnetic field, the separation between the FC and ZFC magnetization is clear below 80 K, which further supports the appearance of magnetic frustrations at low temperatures. The paramagnetic susceptibility in the range $T$ = 250 – 270 K has been fitted to the Curie-Weiss law, $1/\chi = (T - \theta)/C$, [Fig. 2(d)] which gives $\theta$ = 231.42 $\pm$ 2.66 K, $C$ = 7.00 $\pm$ 0.05 emu K/mol Oe and $\mu_{eff}$ = 7.52 $\mu_B$, which is somewhat higher than the theoretically expected value calculated by $\mu_{eff} = \sqrt{\mu_B^2 [g^2 s(s+1)_{Co} + g^2 s(s+1)_{Mn}]} \sim 7.00 \, \mu_B$, where $s$ = 3/2 for both Mn$^{4+}$ and Co$^{2+}$ ions, $g$ = 2 for Mn$^{4+}$ and $g$ = 3 for Co$^{2+}$ due to its orbital contribution [26]. The higher value of $\mu_{eff}$ can imply the presence of FM correlations in the paramagnetic regime, which is consistent with the presence of a FM peak at $T > T_C$ present in the neutron diffraction data, as will be shown in a later section (Fig. 4). The magnetic field dependence of magnetization at selected temperatures [Fig. 3(a)] shows a hysteresis loop signifying ferromagnetism at each temperature except 300 K, at which the material is in the PM state. Magnetization at $H = 50$ kOe and $T = 5$ K is calculated to be $M_{5T} = 5.68$ $\mu_B$/f.u. The structural ordering of B/B′ can be defined as the ratio of experimental to theoretical magnetic moments ($\delta$), which should be 1 for perfect cationic order [27]. In the present case, where theoretical magnetic moment is 6.00 $\mu_B$/f.u., $\delta = 0.95$, which indicates a possible presence of small cationic disorder at B/B′ sites. The coercive field ($H_C$) at $T = 5$ K is $H_C \sim 5.4$ kOe. The remanent magnetization ($M_r$) is 3.10 $\mu_B$/f.u., which indicates a relatively low population of antiphase boundaries, as at $H = 0$ it is expected that any antiphase region would return to its antiparallel orientation, decreasing the value of $M_r$ [6]. Enhanced values of $H_C$ and $M_r$ can also be related to a frozen cluster glass state [28]. Saturation is not achieved even at 50 kOe, which could be due to either large crystalline anisotropy of the octahedral site $Co^{2+}$ ions or canting of antisite spins [6]. The decrease of $M_{5T}$ and $H_C$ with increasing temperature, which is a feature for a typical FM system, is demonstrated in Fig. 3 (b). Based on $H_C \sim 5.4$ kOe at 5 K and the nature of $M$ vs. $H$ curve, it can be concluded that the system is highly anisotropic.

Although the high values of $\mu_{eff}$ and $M_{5T}$ at 5 K relative to previous studies indicate that the sample is highly cationically ordered, [6][13] the value of $\delta < 1$ still suggests the possibility of a mixed valence state of B/B′ site ions, which is mainly due to some concentration of oxygen vacancies as well as possible cation vacancies. The high degree of order of B site cations in the sample implies the presence of predominantly $Mn^{4+}$ and $Co^{2+}$ ions, giving rise to a FM 180° superexchange interaction, $e^2$-O-$e^0$. However, the presence of oxygen vacancies can induce electron transfer from $Co^{2+}$ to $Mn^{4+}$ resulting in intermediate spin $Mn^{3+}$ and $Co^{3+}$ ions leading to vibronic superexchange interactions, $e^1$-O-$e^1$ [6]. Based on the XPS data, which indicates the presence of mixed valence state of Co and Mn ions, and the above discussion, the first FM transition at $T_{F1}$ = 228 K can be attributed to the $Co^{2+}$-O- $Mn^{4+}$ interactions while the second FM transition at $T_{F2}$ = 220 K to that of $Co^{3+}$-O- $Mn^{3+}$ interactions.

## C.    Long-range ferromagnetic ordering

The large separation in FC and ZFC curves and comparatively high values of $H_C$ and $M_r$ seen in DC measurements are indicative of absence of purely FM state in LCMO. To gain insight into the magnetic structure and the order of FM interactions, neutron diffraction experiments were performed. Fig. 4 displays the neutron diffraction patterns of $La_2CoMnO_6$, measured at selected low temperatures. The structural model previously reported for the monoclinic $P2_1/n$ phase provides a satisfactory fit to the room-temperature neutron diffraction data with the fitting reliability R-factors $R_p$ = 6.58% and $R_{wp}$ = 8.35%. Moreover, the large contrast in neutron scattering length of Co (2.49 fm) and Mn (−3.75 fm) allows a determination of their distribution over the crystallographic sites for transition metal ions and the degree of order between the two cations. The cation ordering degree of 0.91(6) was refined, which agrees well with $\delta$ obtained from DC magnetization in Sec. III B. The presence of anion vacancy indicates the existence of mixed valence states of Co ($Co^{2+}/Co^{3+}$) and Mn ($Mn^{3+}/Mn^{4+}$) ions, which, consequently, strongly influence the magnetic properties (Sec. III D). At low temperature, a magnetic contribution to nuclear peaks located at $d_{hkl}$ positions of 2.73 and 3.86 Å is evidenced, which increases continuously as temperature is lowered further. The data analysis reveals that these behaviors correspond to the formation of long-range FM order. The average ordered magnetic moment of Co/Mn ions at 20 K is 2.8(5) $\mu_B$, which is close to the value extracted from isothermal magnetization curves. Notably, the ordered magnetic moment value is lower compared to that determined from spin-only values for Co/Mn ions, which further suggests that the magnetic ground state of La$_2$CoMnO$_6$ is not purely FM.

As the neutron diffraction results indicate predominant long-range ferromagnetic ordering, the critical behavior across the PM-FM phase transition is investigated to further probe the nature of the magnetic ordering. The critical behavior of a second-order phase transition can be characterized by the set of critical exponents, which are defined by universal scaling laws. For a PM-FM transition, the magnetization ($M$) acts as the order parameter, and the universal scaling laws take the form [29][30]:

$$M_S(T) = M_0(-\varepsilon)^\beta, T < T_C \qquad (1)$$

$$\chi^{-1}(T) = (h/M)\varepsilon^\gamma, T > T_C \qquad (2)$$

$$M = DH^{1/\delta}, T = T_C \qquad (3)$$

where $M_0$, $h/M$, and $D$ are the critical amplitudes of the spontaneous magnetization, inverse susceptibility and the field dependence of the magnetization, respectively, and $\varepsilon = (T - T_C)/T_C$ is the reduced temperature. Alternatively, these exponents can be calculated using the Arrott-Noakes equation of state [29][30][31],

$$(H/M)^{1/\gamma} = A\varepsilon + BM^{1/\beta}, \qquad (4)$$

where rescaling of magnetization data as a function of magnetic field and temperature into a series of parallel lines can only be achieved by the correct exponents, and the isotherm corresponding to $T_C$ passes through the origin.

Four different values of the exponents were used to construct the Arrott plot and modified-Arrott plot (MAP) to test the universality class into which this system belongs. For the mean-field model (MF), which is also simply known as the Arrott plot, $\beta = 0.5$ and $\gamma = 1.0$; for 3D Ising, $\beta = 0.325$ and $\gamma = 1.24$; for 3D Heisenberg, $\beta = 0.365$ and $\gamma = 1.336$; and for tricritical mean-field, $\beta = 0.25$ and $\gamma = 1.0$ [32]. Fig. 5(a-d) shows all four plots at high magnetic fields ($\mu_0 H = 2 - 3$ T) and with a step size $\Delta T = 0.25$ K from 220 K – 235 K. The plots show series of quasi-parallel lines with slope given by $S(T) = dM^{1/\beta}/ d(H/M)^{1/\gamma}$. We adopt the approach used in [33] to determine an appropriate model for the system by calculating the normalized slope (NS). Using a value of $T_C = T_{F1}$ determined from $dM/dT$ vs. $T$, NS is defined as NS $= S(T)/ S(T_C)$. NS vs. $T$ is plotted for all the above models and is compared to the ideal value of NS $= 1$. Fig. 6(a) clearly shows that the data rescaled with respect to the mean-field model is closest to 1 while other models deviate significantly. To obtain the most accurate values of the critical exponents and $T_C$, a combined iterative procedure using the Kouvel-Fisher (KF) method [34] and the Arrott-Noakes equation of state was implemented yielding $\beta = 0.59 \pm 0.031$, $\gamma = 1.12 \pm 0.028$ and $T_C = 224 \pm 9.33$ K. The modified-Arrott plot using the values obtained from the KF method is shown in Fig. 6(b). The $\beta$ and $\gamma$ values are closest to MF ($\beta = 0.5$, $\gamma = 1$), in agreement with the normalized slope results, and deviate significantly from the other three models. The exponents indicate that the FM ordering is governed by long-range interactions in the synthesized LCMO. However, the calculated value of $T_C$ differs by 4 K from the value determined by DC magnetization, as well as AC susceptibility (Sec. III D), and has a large error. This discrepancy may be due to the proximity of the onset of the second type of FM correlations at $T_{F2}$ to $T_C$, and as a result, the magnetic system cannot be described as undergoing an exactly canonical PM-FM transition.

## D. Emergent magnetic frustration with short-range order

As shown in Fig. 2(c), the clear separation observed between FC and ZFC curves at fields as high as $\mu_0H = 6$ T signifies the presence of frustration in the system at lower temperatures. In previous studies, glass-like behavior has been reported at low temperatures and associated with possible magnetic frustration. AC magnetic susceptibility is a useful tool to detect the existence of magnetic frustration and characterize associated glassy behavior by an analysis of relaxation phenomena. Furthermore, higher-order susceptibilities can be used to unravel the system's behavior, as they reflect symmetry breaking related to spin configuration [35]. AC susceptibility is the differential $dM/dH$ response of the magnetization of the sample to a time-dependent magnetic field, e.g. $H_{AC} = h$ sin $(2\pi f\ t)$ [35]. The magnetization ($M$) of a system can be expressed in terms of driving field ($H_{AC}$) as:

$$M(H) = M_0 + \chi_1 H + \chi_2 H^2 + \chi_3 H^3 + ... \tag{5}$$

where $M_0$ is the spontaneous magnetization, $\chi_1$ is the linear susceptibility and $\chi_n$ ($n > 1$) are the higher harmonics representing the nonlinear response to $H_{AC}$.

Fig. 7 shows the AC susceptibility measurements at frequencies $f = 500 - 10,000$ Hz. As expected, the real part of first harmonic ($\chi_1'$) (Fig. 7(a)) shows anomalies at $T_{F1} = 228$ K and $T_{F2} = 220$ K, similar to that shown in the ZFC magnetization. For consistency with DC results above, $T_{F1}$ and $T_{F2}$ are considered as the inflection points of the curve. Though the magnitude of the peaks changes with frequency, no longitudinal shift is observed along the temperature axis. In general, for spin glass systems, the longitudinal peak position is dependent on the frequency, as the response of AC-$\chi$ is related to the broad distribution of relaxation times which tend to shift towards longer time scales as the temperature is reduced. The presence of these relaxation times within the time of measurement results in peaks that shift to lower temperature as frequency is decreased [25]. Since the anomalies located at $T_{F1}$ and $T_{F2}$ are frequency *independent*, they can be related to FM interactions of $Co^{2+}$-O-$Mn^{4+}$ and $Co^{3+}$-O-$Mn^{3+}$, respectively.

On closer inspection of Fig. 7(a), two additional features appear as a spread in the curves from $T = 110 - 140$ K and an anomaly at low temperatures, shown in the inset of Fig. 7(a)). In Fig. 7(b), $\chi_1''$ more clearly shows these dynamic features in addition to the peaks at $T_{F1}$ and $T_{F2}$. A broad frequency-*dependent* hump, denoted as $T_{CG1}$, shifts from $T = 110 - 140$ K for frequencies from $f = 500 - 10000$ Hz. The low temperature anomaly observed in $\chi_1'$ appears in the magnetic loss near $T = 42$ K. We denote this transition as $T_{CG2}$. The presence of the frequency-dependent behavior suggests a freezing over a large range of temperatures indicating the existence of a glassy state [36], which is a different observation from the previously reported FM transition near $T = 135$ K [15][19]. To confirm the glass-like behavior, the Vogel-Fulcher (VF) fit has been carried out for the transition at $T_{CG1}$. According to VF model, the relaxation time of an ensemble is given by the following relation,

$$\tau = \tau_0^{\mathrm{VF}} \exp\left[ \mathrm{E}_a \Big/ k_B (T - T_0^{CG1}) \right], \qquad (6)$$

where $E_a$ is the activation energy expressed as the product of anisotropy constant (K) and volume (V), i.e. $E_a = $ KV, $T_0^{CG1}$ is the characteristic temperature, also referred to as the interparticle/intercluster interaction strength, and $\tau_o^{VF}$ is the relaxation time of individual particles. The inset of Fig. 7(b) shows the successful fit to the data. The fitted parameters obtained have the following values: $\tau_o^{VF} = 10^{-7}$ s, $E_a/k_B = 400$ K and $T_0^{CG1} = 77$ K. The relaxation time, $\tau_o^{VF}$, is several orders of magnitude larger than the spin flip time of an atomic magnetic moment ($\sim 10^{-13}$ s) [35] indicating the freezing of spin clusters with a freezing temperature $T_0^{CG1}$

~ 80 K. Since $T_0^{CG1} << E_a/k_B$, there is a weak coupling between the clusters [37]. The Fulcher parameter, $(T_f - T_0^{CG1}) / T_0^{CG1}$, where $T_f$ is the temperature taken as the peak in $\chi''$, is obtained to be 0.68, which is an order of magnitude higher than that of spin glasses, ruling out the possibility of cooperative freezing. Instead, the obtained value is comparable to the case of the progressive freezing of a cluster glass [38]. The peak shift is also characterized by the phenomenological factor which helps to compare glass-like systems, $K = \Delta T_f / (T_f \Delta \log f)$. The calculated value of $K$ for our system is 0.142. The $K$ value has been categorized as 0.005 – 0.01 for spin glass, ~ 0.03 – 0.06 for cluster glass and >0.1 for superparamagnetic compounds [28]. The obtained $K$ value in the superparamagnetic range implies that the intercluster interactions are very weak in the present system.

The ordering of $Mn^{4+}$ and $Co^{2+}$ ions in $La_2Co^{2+}Mn^{4+}O_6$ can be described by FM $e^2$-O-$e^0$ interactions ($T_{F1}$), but appearance of these ions at the antisite positions establishes weaker $Mn^{4+}$-O-$Mn^{4+}$ or $Co^{2+}$-O-$Co^{2+}$ AFM interactions. Dass et al. [6] showed that the formation of antiphase boundaries in an ordered double perovskite leads to the presence of AFM interactions. Antiphase boundaries are formed if the positions of the $Co^{2+}$ and $Mn^{4+}$ ions are inverted in one atomically ordered region relative to that in a neighboring region. The resulting antiphase interface between the two regions would have short range AFM $Co^{2+}$-O-$Co^{2+}$ or $Mn^{4+}$-O- $Mn^{4+}$ interactions while the ordered regions would have FM $Co^{2+}$-O-$Mn^{4+}$ interactions. Thus, the observed spin clusters are likely to arise due to the local magnetic frustration caused by the competition between the FM and AFM interactions.

With similar arguments as above, the low temperature kink at $T_{CG2}$ = 42 K seen in Fig. 7(b), can be associated with oxygen deficiency. The oxygen-deficient regions consist of $Co^{3+}$ and $Mn^{3+}$ ions which interact via $Co^{3+}$-O-$Mn^{3+}$ FM vibronic superexchange interactions. The presence of antisite disorder in these regions would lead to short-range AFM $Co^{3+}$-O-$Co^{3+}$ or $Mn^{3+}$-O-$Mn^{3+}$ interactions at the antiphase interface. Thus, the observed anomaly at $T_{CG2}$ could be due to the local frustrations arising from the competitive $Co^{3+}$-O-$Mn^{3+}$ FM and $Co^{3+}$-O-$Co^{3+}$ or $Mn^{3+}$-O-$Mn^{3+}$ AFM interactions. The absence of a frequency-dependent feature at $T_{CG2}$ could be due to the longer time scales required to observe the dynamics of these clusters at low temperature.

To review, the presence of antisite defects in regions with and without oxygen vacancies lead to the formation of two types of clusters in the system, respectively. Type I: co-existence of FM $Co^{2+}$-O-$Mn^{4+}$ and short range AFM $Co^{2+}$-O-$Co^{2+}$ or $Mn^{4+}$-O-$Mn^{4+}$ interactions and type II: co-existence of FM $Co^{3+}$-O-$Mn^{3+}$ and AFM $Co^{3+}$-O-$Co^{3+}$ or $Mn^{3+}$-O-$Mn^{3+}$ interactions. It is expected that the volume fraction of the anion-deficient region is low, which would result in a smaller volume of type II clusters compared to type I. For a certain field and frequency, the dynamics of larger clusters freeze at a higher temperature. Hence, the freezing of type I clusters is observed before type II clusters.

To further support the FM nature of the transition at $T_{F1}$, the nonlinear components of the AC susceptibility were studied. The real part of second harmonic component ($\chi_2'$) (Fig. 7(c)) is zero in the paramagnetic phase, has a positive peak at 228 K ($T_{F1}$), a negative peak at 218 K (~$T_{F2}$) and thereafter slowly decreases to a small yet finite value at the lowest temperatures measured. The strength of the positive peak at $T_{F1}$ weakens with an increase in the frequency. According to Pramanik et. al. [39], $\chi_2$ can be observed experimentally only if there is presence of a symmetry breaking internal field, or spontaneous magnetization, i.e. $\Delta M = M(H) - (-M(-H)) \neq 0$. Sudden changes in the internal field result in sharp features in $\chi_2$ [39]. Hence the sharp peak in $\chi_2'$ [Fig. 7(c)] corresponds to FM $Co^{2+}$-O-$Mn^{4+}$ interactions, and the sharp change in peak direction near $T_{F2}$ indicates a sudden change in the internal field. This reinforces that a modification of the magnetic structure occurs at $T_{F2}$, where the FM interactions of $Co^{3+}$ and $Mn^{3+}$ ions are stabilized. At low temperatures, a small yet finite value of $\chi_2'$ remains indicating the presence of spontaneous magnetization, which, however, lacks any indication of sudden variations in the magnetic ordering.

The real part of third harmonic ($\chi_3'$) is shown in Fig. 7(d). The anomaly in $\chi_3'$ is known to indicate the presence of a magnetic phase change, and reflects the nature of the transition at $T_C$ [35]. The negative to positive crossover of $\chi_3'$ at $T = 228$ K indicates the paramagnetic to ferromagnetic nature of the phase transition, hence this temperature is defined as $T_C$ for this system. Furthermore, the inflection point of $\chi_1'$, as shown in the inset of Fig. 7(d), and the inflection points of the ZFC and FC curves [Fig. 2(b)] justifies the definition of $T_C = 228$ K ($T_{F1}$).

To gain a better understanding of the effects of the coexisting cluster glass state and ferromagnetic phase on the magnetic properties, the magnetic anisotropy is investigated as a function of temperature. While the static magnetization curves can give a rough estimation of the total anisotropy, transverse susceptibility measurements provide a much more accurate determination [20]. Transverse susceptibility (TS) measurements are performed using a custom built tunnel diode oscillator (TDO) probe, [20] which provides basic information about how the magnetic anisotropy evolves with temperature. The change in resonant frequency ($\Delta f$) of the TDO circuit is a consequence of the change in inductance as the sample inside the circuit is magnetized. $\Delta f$ is directly proportional to the change in TS ($\Delta \chi_T$) such that the quantity

$$\frac{\Delta \chi_T}{\chi_T}(\%) = \frac{|\chi_T(H) - \chi_T^{sat}|}{\chi_T^{sat}} x100 \qquad (7)$$

can be measured as a function of $H_{DC}$, where $\chi_{T}{}^{sat}$ is the TS at the saturating or maximum field, $H_{sat}$. Peaks are expected in the TS scan at the positive and negative anisotropy fields, $\pm H_K$, and at the switching fields, $\pm H_S$. As the TS measured via TDO is a dynamic response in nature, it provides an ideal route to probe the effects of the time-dependent properties of the coexisting cluster glass and FM phase on $H_K$.

Fig. 8(a) shows the TS measurement at 20 K. $\pm H_K$ are shown by arrows. The switching field is generally merged with one of these peaks [40]. The broad nature of the peaks could be due to the distribution of the anisotropy axes in the polycrystalline sample [20]. Fig. 8(b) shows the dependence of $H_K$ on temperature which displays a maximum at $T \sim 80$ K. Above 80 K, the reduction in $H_K$ with temperature behaves as expected in a FM system, where thermal fluctuations tend to lower the energy barrier that the external field must overcome to align spins against their anisotropy axis. However, below ~80 K, $H_K$ values drop continuously with decreasing temperature. This observation is in stark contrast to the behavior observed in the magnetic hysteresis loops; Since the coercivity increases as temperature is lowered [Fig. 3 (b)], it is expected that the total magnetic anisotropy of the system would show a similar trend. The difference in the temperature dependence of $H_K$ and $H_C$ suggests that the TS measurement does not reflect the total anisotropy of the magnetic system over the entire temperature range. Moreover, the maximum in $H_K$ at $T = 80$ K agrees well with the freezing temperature of $T_0{}^{CG1} = 77$ K determined for the type I clusters in the AC-$\chi$ measurements. $H_K$ begins to decrease as temperature drops below $T_0{}^{CG1}$. This behavior may signify the pinning of clusters to the FM matrix, which effectively reduces the contribution of the random anisotropy of the clusters reflected in the dynamic response. The continuous decrease in $H_K$ as temperature is reduced further reflects the progressive freezing of spin clusters in the type II regime. It should be noted that TDO measurements are performed at relatively high frequencies such that the behavior associated with the dynamics of the FM matrix + clusters can be markedly different from the behavior observed in the zero frequency $M$ vs. $H$ measurements.

### E. Magnetocaloric effect and magnetic phase diagram

With the understanding gained about the mechanisms of the phase evolution using time-dependent measurements, a detailed investigation of the static magnetic behavior across the $H$-$T$ phase diagram is performed by exploiting the magnetocaloric effect. On subjecting the magnetic sample to a change of external magnetic field at a constant temperature, the isothermal entropy change, or by convention, the magnetic entropy change ($\Delta S_M$), can be calculated with the help of the following Maxwell relation,

$$\left(\frac{\partial S_M(T,H)}{\partial H}\right)_T = \mu_0 \left(\frac{\partial M(T,H)}{\partial T}\right)_H .  \qquad (8)$$

The magnetic entropy change may be calculated numerically via

$$\Delta S_M = \mu_0 \int_{H_i}^{H_f} \left(\frac{\partial M}{\partial T}\right) dH' \qquad (9)$$

Conventionally, in ferromagnetic materials, the magnetic entropy of the spin system decreases ($\Delta S_M < 0$) as the magnetic field tends to orient moments along the field direction, hence suppressing thermal fluctuations. On the other hand, the application of a magnetic field may increase the entropy causing $\Delta S_M > 0$, for example in antiferromagnets, as the external field rotates the spins in antiparallel sublattices against their preferred direction [29][41][42]. The magnetic entropy change has been demonstrated to be an effective tool to understand the features of the coexisting magnetic phases [29][41][42][43]. In the following discussion, the magnetic entropy change is analyzed as a function of temperature and magnetic field across the *H-T* phase diagram.

Based on isothermal magnetization versus magnetic field curves, the calculated magnetic entropy change as a function of temperature and magnetic field change, $\mu_0\Delta H$, is summarized in Fig. 9. The plot of $\Delta S_M$ (*T*) at low changes in field, $\mu_0\Delta H = 0.05 - 0.45$ T, is shown in Fig. 9(a). The dominant feature in $\Delta S_M$ (*T*) is the minimum near $T_{F1} = T_C$ that signifies the transition from the PM phase into the long-range-ordered FM state. Fig. 9(b) displays a zoomed view of $\Delta S_M$ (*T*) at high fields and high temperature in which, along with the negative peak at $T_{F1}$, a dip in magnetic entropy occurs at $T_{F2}$ (red dashed line). The observation of this dip in entropy at $T_{F2}$ corroborates our earlier assignment of this peak to the onset of FM $Co^{3+}$-O-$Mn^{3+}$ interactions. At low temperatures, however, a field-dependent crossover from negative to positive $\Delta S_M$ (*T*) is observed in Fig. 9(a), which signifies a field-induced disordering of the magnetic state. Fig. 9(c) displays the magnetic field dependence of $\Delta S_M$ at low temperatures, which reaches a maximum positive value before decreasing at higher $\mu_0\Delta H$.

We first analyze the magnetic entropy change across the PM-FM phase transition. Based on power law dependence of change in magnetic entropy with field, i.e. $\Delta S_M \propto H^n$, it has been confirmed that $\Delta S_M$ (*T*) can be scaled close to the second-order phase transition [29][44]. Hence if proper scaling is achieved, $\Delta S_M$ (*T*) curves close to the transition temperature and corresponding to different fields should collapse on to the same point of the universal curve. According to the procedure described in [45], the universal curve is constructed as follows: $\Delta S_M$ (*T*) is normalized by the maximum value of $\Delta S_M$ ($T_{peak}$), where $T_{peak}$ is the transition temperature, the temperature axis is rescaled such that $\Delta S_M (T_r)/ \Delta S_M (T_{peak}) \geq 0.5$, where $T_r$ is a reference temperature. The rescaled temperature axis is defined as [45]

$$\theta = \begin{cases} -\dfrac{T - T_C}{T_{r1} - T_C} & T \leq T_C \\[2mm] -\dfrac{T - T_C}{T_{r2} - T_C} & T > T_C \end{cases} , \qquad (10)$$

where $T_{r1}$ and $T_{r2}$ are the two reference temperatures, $T_{r1} > T_{peak}$ and $T_{r2} < T_{peak}$ and are chosen such that $\Delta S_M (T_{r1})/ \Delta S_M (T_{peak}) = \Delta S_M (T_{r2})/ \Delta S_M (T_{peak}) = 0.70$.

Fig. 9(d) shows the universal curve constructed for field range of $0.14 - 0.75$ T, which provides evidence for the second-order PM-FM phase transition and corroborates the critical analysis performed in Sec. III B. However, below $T_{peak}$ a spread in the data starts from ~ 211 K with a dispersion of ~16%. Failure of collapse for $\theta < $ -1 has been ascribed to improper scaling of magnetic entropy and may be attributed to the increasing fluctuations of an additional magnetic phase near the ordering temperature, $T_{peak}$ [29][46]. This failure can be eliminated by the use of two reference temperatures, and collapse can be achieved for a second-order transition. An absence of universal behavior even with the use of two reference temperatures is typically associated with the presence of a first-order transition and typically exceeds 100% [29][44][46]. In our results, however, a significant dispersion is observed above $T_{r2}$ ($\theta > $ -1). This region lies in the temperature range at which the $Co^{3+}$-O-$Mn^{3+}$ FM interactions enter the system just below $T_C$. Thus, the occurrence of dispersion is likely due to the proximity of the critical temperature for the onset of second FM interactions to $T_{peak}$. The inset of Fig. 9(d) shows the rescaling of field axis to produce a linear plot of $\Delta S_M (T_{peak})$ vs. $H^n$, where $n = 1 + (1 - \beta)/(\beta + \gamma) = 2/3$ for mean-field [30][45]. The linearity of graph further supports the observation of mean-field critical behavior at the phase transition, as concluded in Sec. III C.

The *H-T* phase diagram in Fig. 10 is formulated based on the results from static and dynamics measurements. Namely, it displays a surface plot that illustrates the general behavior of $\Delta S_M$ due to the changes in field and temperature, where cool colors signify the region of decreasing $\Delta S_M$ and warm colors indicate the region of increasing $\Delta S_M$. The minima [Fig. 9(b)] corresponding to $T_{F1}$, where universal behavior is observed, and $T_{F2}$ are marked by yellow and red dashed lines, respectively. Characteristic temperatures derived from AC measurements are displayed: The red star labels the freezing temperature, $T_0^{CG1}$, where the dynamics of type I clusters become frozen, and the blue star marks the type II cluster anomaly seen in $\chi_1$.

As temperature is lowered into the cluster glass regime, the region of positive $\Delta S_M$ broadens rapidly and begins to dominate the *H-T* phase diagram. The presence of non-negative $\Delta S_M$ $(T, \mu_0 \Delta H)$ is likely due to the magnetic field orienting the spins within each cluster away from their preferred direction, i.e. lowest energy configuration, [29] or against the local anisotropy within the clusters [47]. Although ND results indicate predominantly long-range ferromagnetic order for temperatures as low as 20 K, the glassy states arising due to the competing short-range AFM interactions become stabilized at sufficiently low thermal energy and dominate the magnetic entropy change. As seen in Fig. 9(b), $\Delta S_M$ reaches a maximum after which it begins to decrease with magnetic field. This indicates a crossover at which the external field is sufficient to overcome the random local anisotropy of the individual clusters, as shown by the black line in Fig. 10. A jump in the critical magnetic field is observed at $T_0^{CG1}$, where $H_K$ begins to decrease. The glassy phases are separated by a dashed line, where the red star represents the freezing temperature of type I clusters. At temperatures below the dashed line, type II clusters progressively freeze, as demonstrated in Sec. III D. A rapid increase in the critical field occurs as temperature is lowered, especially near $T_{CG2}$ (blue star). In this regime, the progressive freezing phenomena of type II clusters shows up as a dramatic increase in magnetic entropy and the peak-like behavior of $\Delta S_M$ ($\mu_0 \Delta H$) broadens significantly. At fields up to 2 T, the entropy change has not crossed to negative, as seen in Fig. 9(b). However, according to equation (9), $dM/dT < 0$ at $\mu_0 H = 6$ T [Fig. 2(c)] implies that even for $T < 50$ K, $\Delta S_M$ ($T$) will become negative at adequately high magnetic field.

## IV. CONCLUSION

A comprehensive study of the static and dynamic magnetic behavior of the double perovskite, La$_2$CoMnO$_6$ (LCMO), has been performed to clarify the complex phase evolution. A long range ferromagnetic order is established at $T_{F1} = 228$ K. Unlike previous studies, we found the high values of $T_{F1}$ and $\mu_{eff}$ in the sample with a relatively large degree of cationic order and noticeable amount of oxygen deficiency, the latter leading to the second FM interaction established at 220 K. The presence of antisite disorder and oxygen deficiency in a long-range-ordered FM matrix serves as the origin of two different types of clusters. The dynamics of type I clusters are clearly seen in the AC-$\chi$ data, while type II clusters relax at longer time scales. Using transverse susceptibility to study the behavior of the magnetic anisotropy as a function of temperature, the freezing mechanism of the cluster glass-like states was investigated. The behavior of the magnetic entropy change supports the existence of a cluster-glass regime, as the increasing entropy indicates that the spin configuration is disordered when the external field is applied. Magnetic entropy change results clearly show the coexistence of two different FM phases and the spin cluster dominance at low temperatures. A comprehensive magnetic phase diagram was constructed which summarizes the static and dynamic features of the magnetic phases.

**ACKNOWLEDGEMENTS**

Research at USF (synthesis and magnetic study) was supported by the U.S. Department of Energy, Office of Basic Energy Sciences, Division of Materials Sciences and Engineering under Award No. DE-FG02-07ER46438.   Work at DTU was supported by the Vietnam National Foundation for Science and Technology Development (NAFOSTED) under grant number 103.02-2017.364.

**REFERENCES**

[1]    S. Vasala and M. Karppinen, Prog. Solid State Chem., **43**, 1–36, (2015).

[2]    G. King and P. M. Woodward, J. Mater. Chem., **20**, 5785, (2010).

[3]    M. T. Anderson, K. B. Greenwood, G. A. Taylor, and K. R. Poeppelmeier, Prog. Solid State Chem., **22**, 197–233, (1993).

[4]    R. N. Mahato, K. Sethupathi, and V. Sankaranarayanan, J. Appl. Phys., **107**, 09D714, (2010).

[5]    M. P. Singh, S. Charpentier, K. D. Truong, and P. Fournier, Appl. Phys. Lett., **90**, 1–4, (2007).

[6]    R. I. Dass and J. B. Goodenough, Phys. Rev. B, **67**, 014401, (2003).

[7]    Y. Mao, J. Parsons, and J. S. McCloy, Nanoscale, **5**, 4720, (2013).

[8]    N. S. Rogado, J. Li, A. W. Sleight, and M. A. Subramanian, Adv. Mater., **17**, 2225–2227, (2005).

[9]    S. Zhao, L. Shi, S. Zhou, J. Zhao, H. Yang, and Y. Guo, J. Appl. Phys., **106**, 123901, (2009).

[10]   H. Guo, J. Burgess, S. Street, A. Gupta, T. G. Calvarese, and M. A. Subramanian, Appl. Phys. Lett., **89**, 022509, (2006).

[11]   K. D. Truong, J. Laverdière, M. P. Singh, S. Jandl, and P. Fournier, Phys. Rev. B, **76**, 132413, (2007).

[12]    M. P. Singh, K. D. Truong, and P. Fournier, Appl. Phys. Lett., **91**, 042504, (2007).

[13]    John B. Goodenough, John Wiley and Sons, New York-London, (1963) .

[14]    S. Yáñez-Vilar, M. Sánchez-Andújar, J. Rivas, and M. A. Señarís-Rodríguez, J. Alloys Compd., **485**, 82–87, (2009).

[15]    H. Z. Guo, A. Gupta, J. Zhang, M. Varela, and S. J. Pennycook, Appl. Phys. Lett., **91**, 202509, (2007).

[16]    M. Balli, P. Fournier, S. Jandl, K. D. Truong, and M. M. Gospodinov, J. Appl. Phys., **116**, 073907, 2014.

[17]    A. J. Barón-González, C. Frontera, J. L. García-Muñoz, B. Rivas-Murias, and J. Blasco, J. Phys. Condens. Matter, **23**, 496003, (2011).

[18]    X. L. Wang, M. James, J. Horvat, and S. X. Dou, Supercond. Sci. Technol., **15**, 427–430, (2002).

[19]    J. Krishna Murthy and A. Venimadhav, J. Appl. Phys., **113**, 1–6, (2013).

[20]    N.A. Frey Huls, N.S. Bingham, M. H. Phan, H. Srikanth, D. D. Stauffer and C. Leighton, Phys. Rev. B, **83**, 024406, (2011).

[21]    V.L. Aksenov, A.M. Balagurov, V.P. Glazkov, D.P. Kozlenko, I.V. Naumov, B.N. Savenko, D.V. Sheptyakov, V.A. Somenkov, A.P. Bulkin, V.A. Kudryashev, and V.A. Trounov, Physica B **265,** 258 (1999)

[22]    J. Rodríguez-Carvajal, Physica B **192**, 55 (1993)

[23]    Stavitski E, Frank M. F. de Groot, Micron, **41**, 687-694, (2010).

[24] V. Cuartero, S. Lafuerza, M. Rovezzi, J. García, J. Blasco, G. Subías, and E. Jiménez, Phys. Rev. B, **94**, 155117, (2016).

[25] J. A. Mydosh, Hyperfine Interact., **31**, 347–362, (1986).

[26] A. N. Vasiliev, O. S. Volkova, L. S. Lobanovskii, I. O. Troyanchuk, Z. Hu, L. H. Tjeng, D. I. Khomskii, H.-J. Lin, C. T. Chen, N. Tristan, F. Kretzschmar, R. Klingeler, and B. Büchner, Phys. Rev. B, **77**, 104442, (2008).

[27] G. Blasse, J. Phys. Chem. solids, **26**, 1969-1971,(1965).

[28] J. A. Mydosh, Taylor and Francis, London, (1993).

[29] Eleanor M. Clements, Raja Das, Ling Li, Paula J. Lampen-Kelley, Manh-Huong Phan, Veerle Keppens, David Mandrus and Hariharan Srikanth, Sci. Rep., **7**, 6545, (2017).

[30] P. Lampen, M. H. Phan, H. Srikanth, K. Kovnir, P. Chai, and M. Shatruk, Phys. Rev. B, **90**, 174404, (2014).

[31] A. Arrott and J. E. Noakes, Phys. Rev. Lett., **19**, 786, (1967).

[32] S. N. Kaul, J. Magn. Magn. Mater., **53**, 5–53, (1985).

[33] Lei Zhang, Hui Han, Min Ge, Haifeng Du, Chiming Jin, Wensen Wei, Jiyu Fan, Changjin Zhang, Li Pi, and Yuheng Zhang, Sci. Rep., **6**, 22397, (2016).

[34] J. S. Kouvel and M. E. Fisher, Phys. Rev., **136**, A1626-A1632, (1964)

[35]  M. Bałanda, Acta Phys. Pol. A, **124**, 964–976, (2013).

[36] C. A. M. Mulder, A. J. van Duyneveldt, and J. A. Mydosh, Phys. Rev. B, **23**, 1384–1396, (1981).

[37]   S. Shtrikman and E. P. Wohlfarth, Phys. Lett. A, **85** , 467-470, (1981).

[38]   J. L. Dormann, D. Fiorani and E. Tronc, J. Magn. Magn. Mater., **202**, 251-267, (1999).

[39]   A. K. Pramanik and A. Banerjee, J. Phys. Condens. Matter, **20**, 275207, (2008).

[40]   P. Poddar, J. Wilson and H. Srikanth, Phys. Rev. B, **68**, 214409, (2003).

[41]   M.-H. Phan and S.-C. Yu, J. Magn. Magn. Mater., **308**, 325–340, (2007).

[42]   P. Lampen, N. S. Bingham, M. H. Phan, H. Srikanth, H. T. Yi, and S. W. Cheong, Phys. Rev. B, **89**, 144414, (2014).

[43]   N. S. Bingham, A. K. Suszka, C. A. F. Vaz, H. Kim, and L. J. Heyderman, Phys. Rev. B, **96**, 024419, (2017).

[44]   C. M. Bonilla, J. Herrero-Albillos, F. Bartolomé, L. M. García, M. Parra-Borderías, and V. Franco, Phys. Rev. B, **81**, 224424, (2010).

[45]   V. Franco, A. Conde, J. M. Romero-Enrique, and J. S. Blázquez, J. Phys. Condens. Matter, **20**, 285207, (2008).

[46]   V. Franco, R. Caballero-Flores, A. Conde, Q. Y. Dong, and H. W. Zhang, J. Magn. Magn. Mater., **321**, 1115–1120, (2009).

[47]   Anis Biswas, Sayan Chandra, Tapas Samanta, Barnali Ghosh, Subarna Datta, M. H. Phan, A. K. Raychaudhuri, I. Dass and H. Srikanth, Phys. Rev. B, **87**, 134420, (2013).

**FIGURE CAPTIONS**

**FIG. 1** Characterization of structural and cationic ordering. (a) XRD pattern, (b)SEM images of the single phase polycrystalline LCMO sample, and XPS spectra at the (c) Mn and (d) Co $L_{2,3}$ edges.

**FIG. 2** Temperature dependence of the magnetization, $M$ vs. $T$, (a) under field-cooled (FC) and zero field-cooled (ZFC) protocols. (b) $dM/dT$ for FC and ZFC data. (c) FC and ZFC $M$ vs. $T$ at $H$ = 6 T. (d) The Curie-Weiss fit of susceptibility in the paramagnetic phase.

**FIG. 3** Magnetization versus magnetic field, $M$ vs. $H$, (a) at different temperatures and (b) the temperature dependence of coercive field ($H_c$) and magnetization at $H$ = 5 T ($M_{5T}$).

**FIG. 4** The neutron diffraction patterns of $La_2CoMnO_6$, measured at low temperatures and processed by the Rietveld method. The experimental points and calculated profiles are shown. Ticks below represent calculated positions of the nuclear peaks of the monoclinic $P2_1/n$ phase. The peaks with ferromagnetic contribution are denoted by symbol "FM".

**FIG. 5** The modified-Arrott plots, $M^{1/\beta}$ vs. $(H/M)^{1/\gamma}$, using (a) mean-field, (b) 3D-Ising, (c) 3D-Heisenberg and (d) tricritical mean-field exponents.

**FIG. 6** Determination of the critical exponents using (a) normalized slope versus temperature (a) and (b) the modified-Arrott plot using $\beta$ and $\gamma$ obtained from the KF method.

**FIG. 7** Temperature dependence of the AC susceptibility. (a) The real part, $\chi_1'$, and (b) imaginary part, $\chi_1''$, of the linear susceptibility. The real parts of the (c) second harmonic, $\chi_2'$, and (d) third harmonic, $\chi_3'$, of the AC susceptibility with an amplitude $h$ = 5 Oe at various

frequencies. Inset (a) shows the zoomed view of $\chi_1'$ from $T = 40 - 80$ K. Inset (b) shows the Vogel-Fulcher fit for frequency dependent peaks, $T_{CG1}$. Inset (d) shows $d\chi_1'/dT$ vs. $T$.

**FIG. 8** Determination of the temperature dependence of magnetic anisotropy using transverse susceptibility. (a) Bipolar TS scans as a function of applied magnetic field for $T = 20$ K. The peaks in TS denote the magnetic anisotropy fields, $H_K$. (b) Dependence of $H_K$ on temperature. The red line represents a Gaussian fit to the data.

**FIG. 9** The magnetic entropy change ($\Delta S_M$) (a) as a function of temperature at low applied magnetic fields and (b) at high field and high temperature. (c) Magnetic field dependence of $\Delta S_M$ for low temperatures in the vicinity of $T_{CG2}$. (d) Universal curve of the magnetic entropy change, $\Delta S_M / \Delta S_M^{max}$ vs. $\theta$.

**FIG. 10** $H$-$T$ phase diagram of single phase $P2_1/n$ La$_2$CoMnO$_6$. The surface plot shows $\Delta S_M$ over the full range of temperatures and changes in magnetic field studied. $T_{F1} = T_C$ (dashed yellow line) marks the PM-FM phase transition corresponding to Co$^{2+}$-O-Mn$^{4+}$ 180º superexchange interactions, $T_{F2}$ (dashed red line) identifies the onset of Co$^{3+}$-O-Mn$^{3+}$ FM interactions attributed to vibronic superexchange. Regions of positive $\Delta S_M$ indicate that the magnetic field disorders the cluster glass state, which is increasingly stabilized as temperature decreases. $T_0^{CG1} = 77$ K (red star) marks the freezing temperature of type I clusters, determined using the Vogel-Fulcher model, and $T_{CG2} = 42$ K (blue star) marks the anomaly in AC-$\chi_1$ associated with type II clusters. The black line indicates the temperature dependence of the crossover field above which positive $\Delta S_M$ begins to decrease. Jumps in the crossover field occur at the characteristic temperatures of type I and type II clusters, $T_0^{CG1}$ and $T_{CG2}$, respectively.

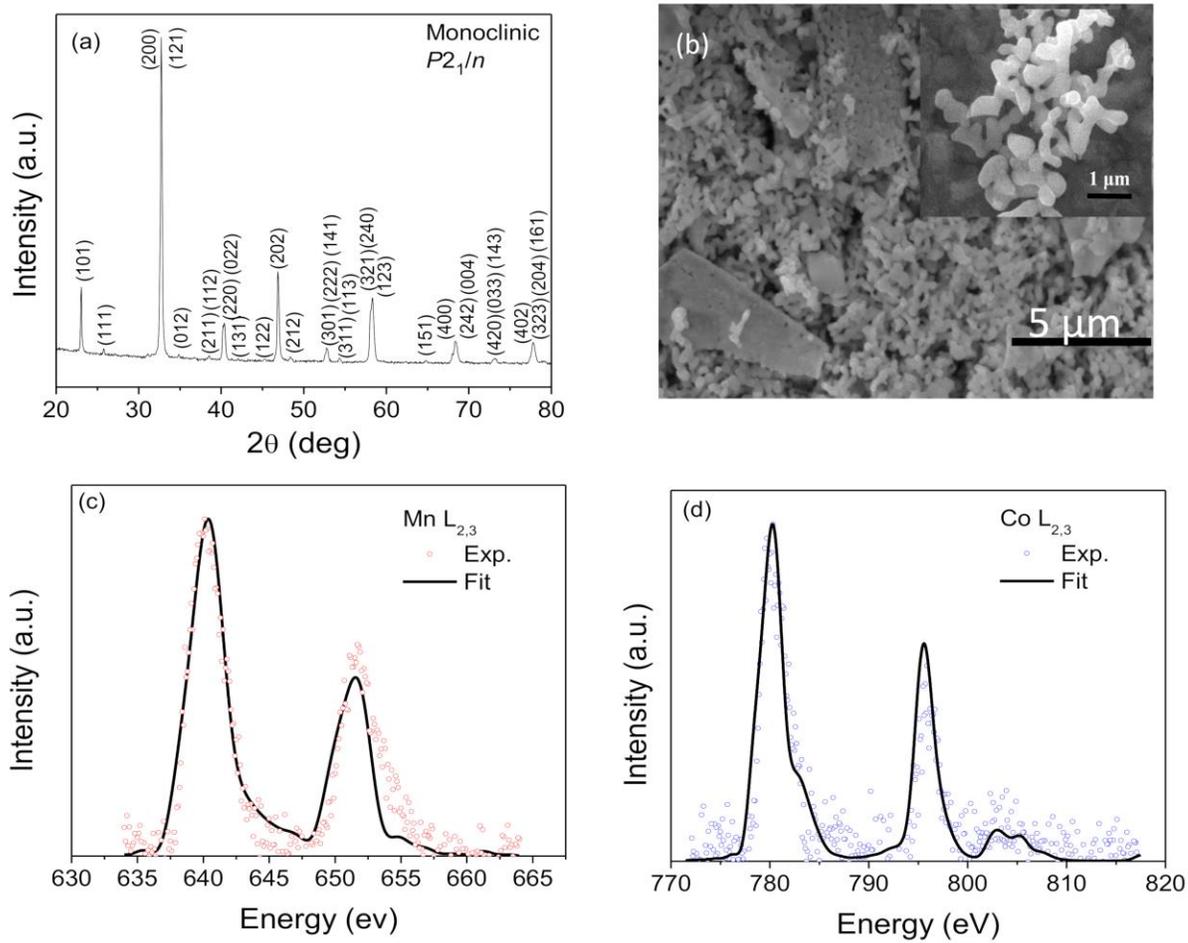

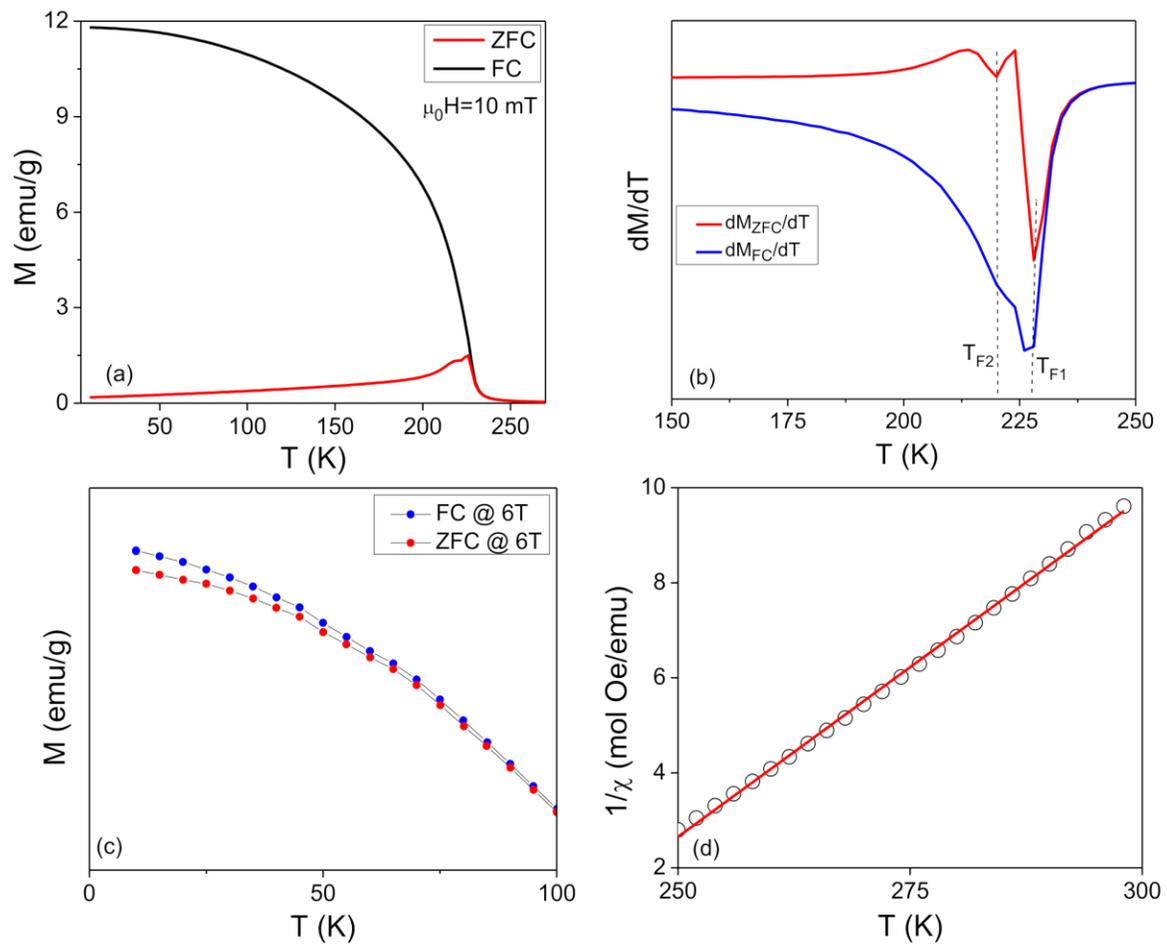

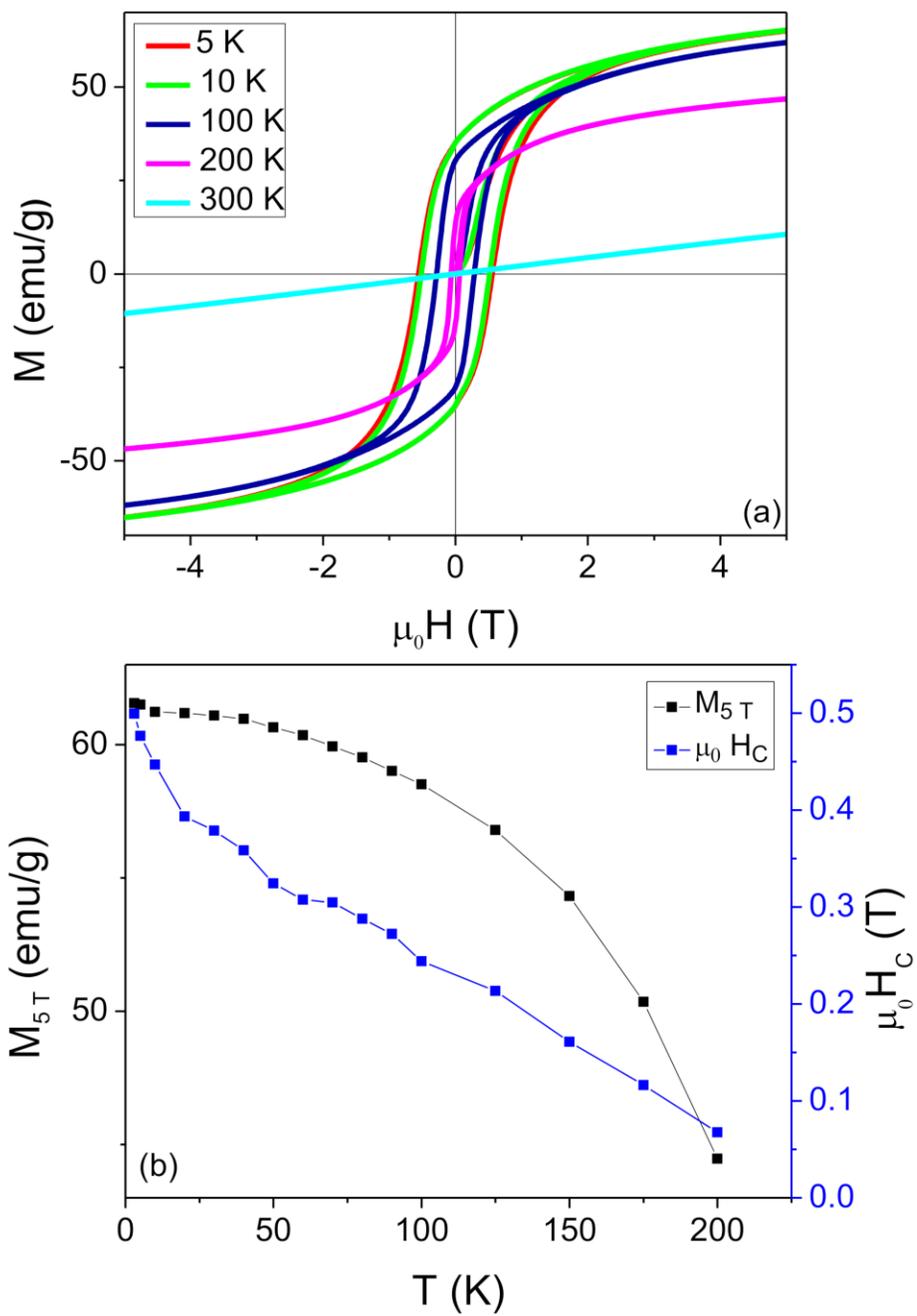

**FIG. 4**

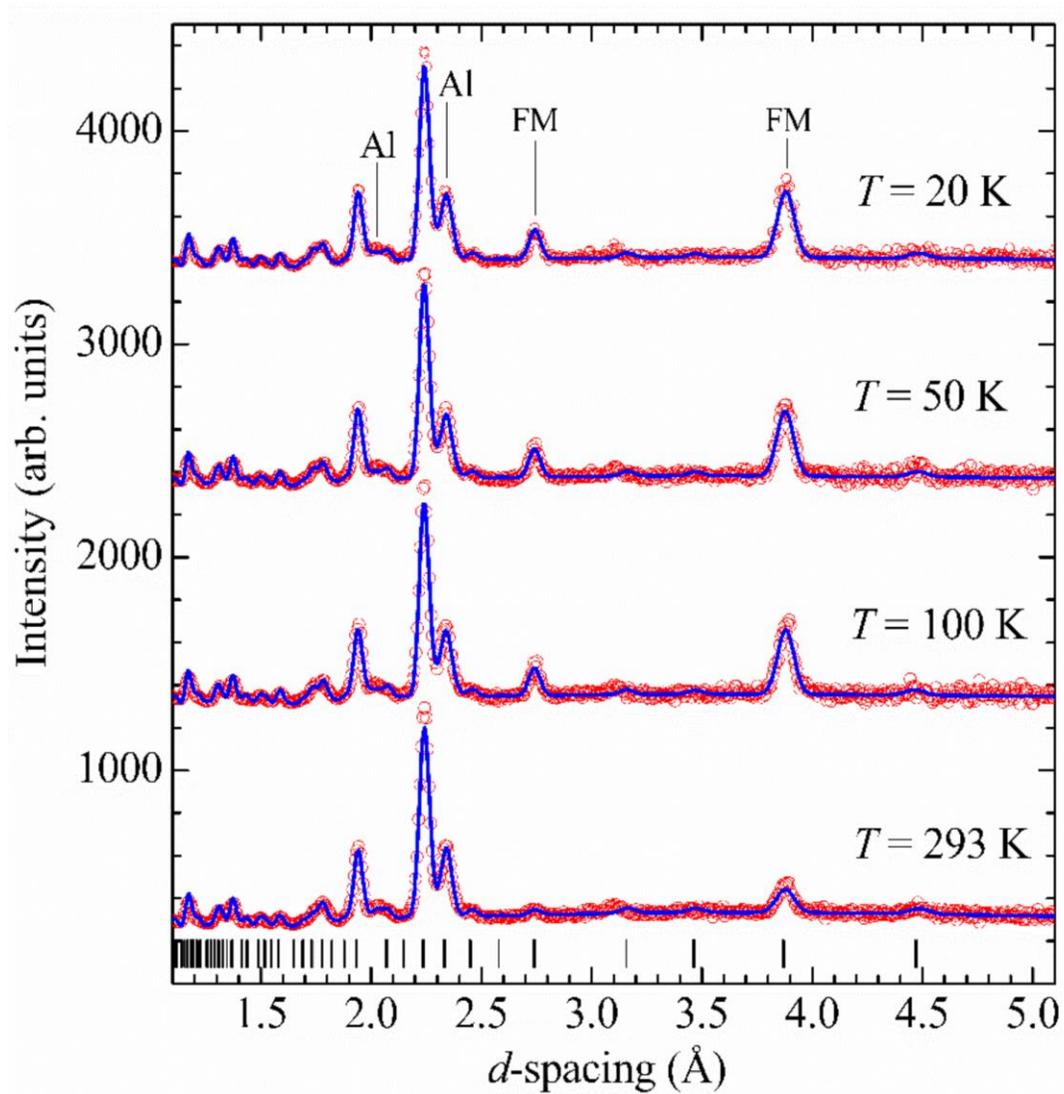

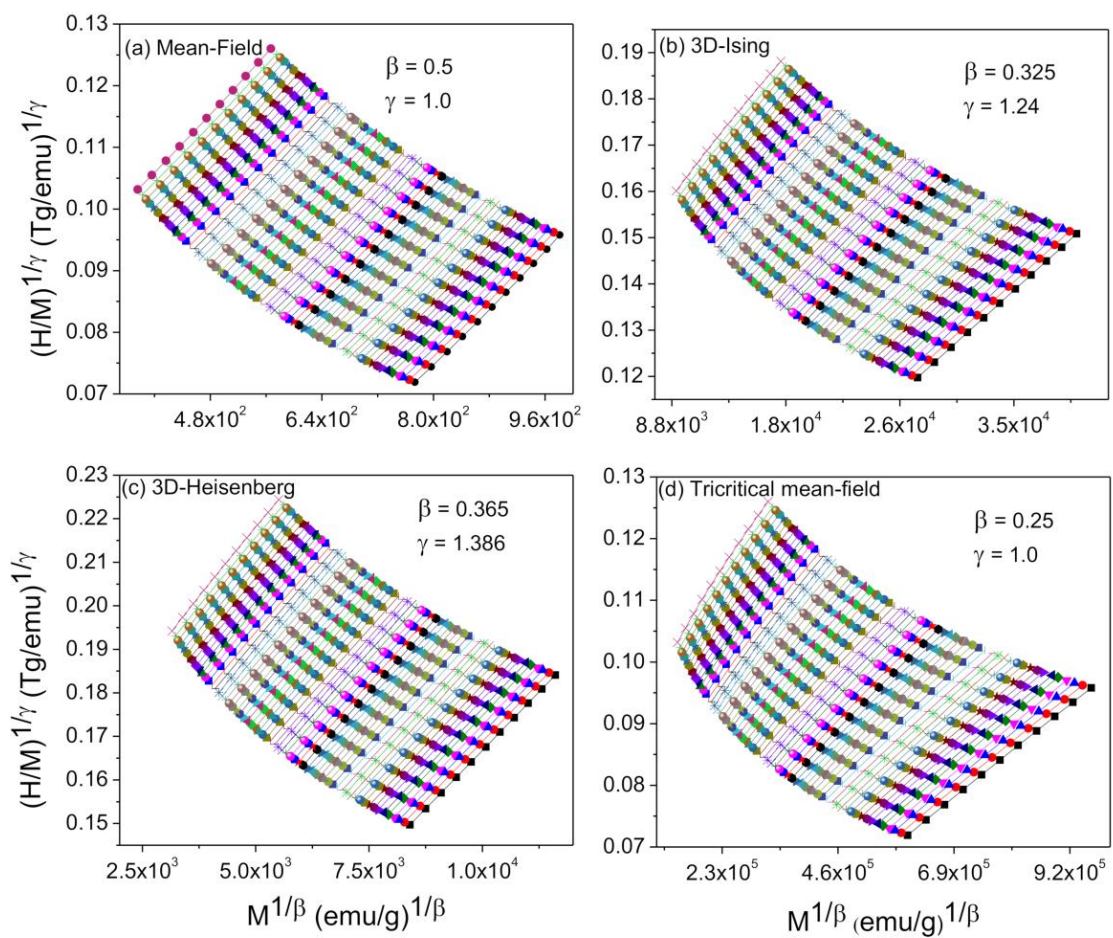

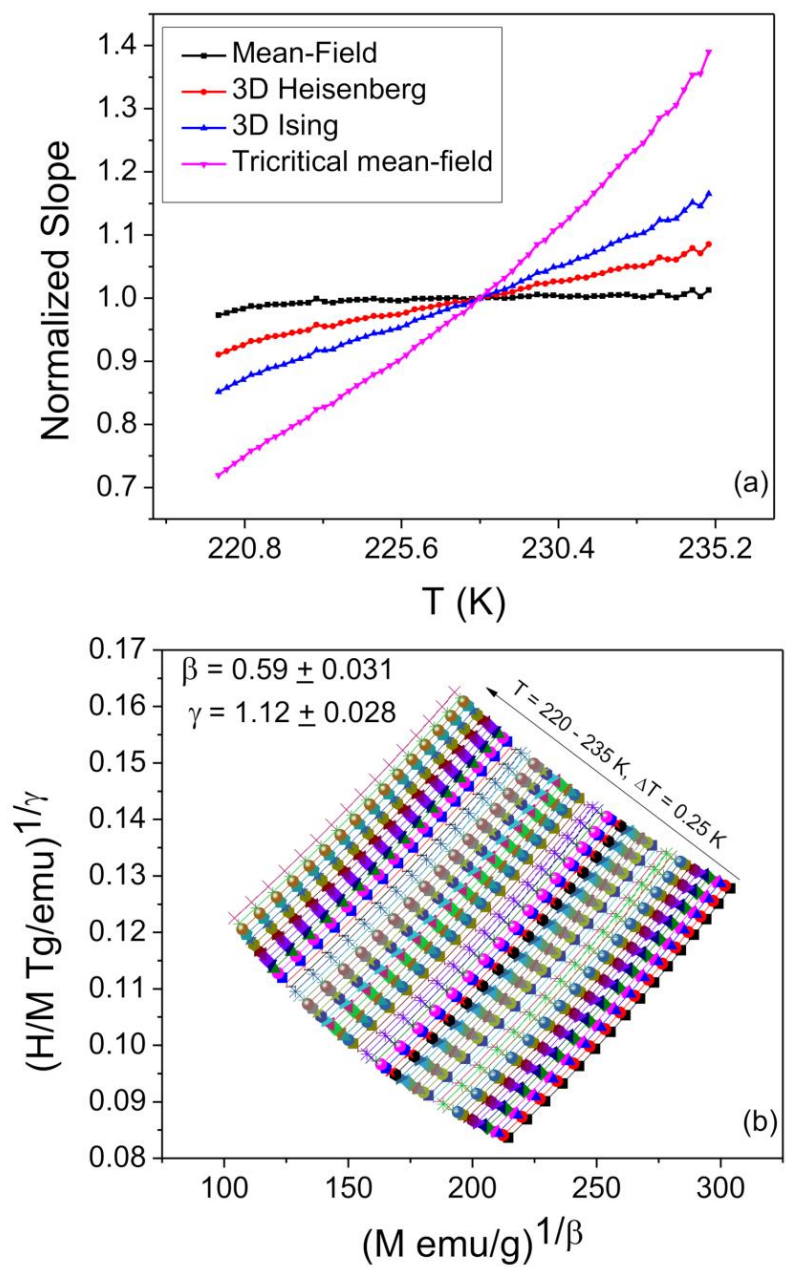

(a)

(b)

$\beta = 0.59 \pm 0.031$
$\gamma = 1.12 \pm 0.028$

$T = 220 - 235$ K, $\Delta T = 0.25$ K

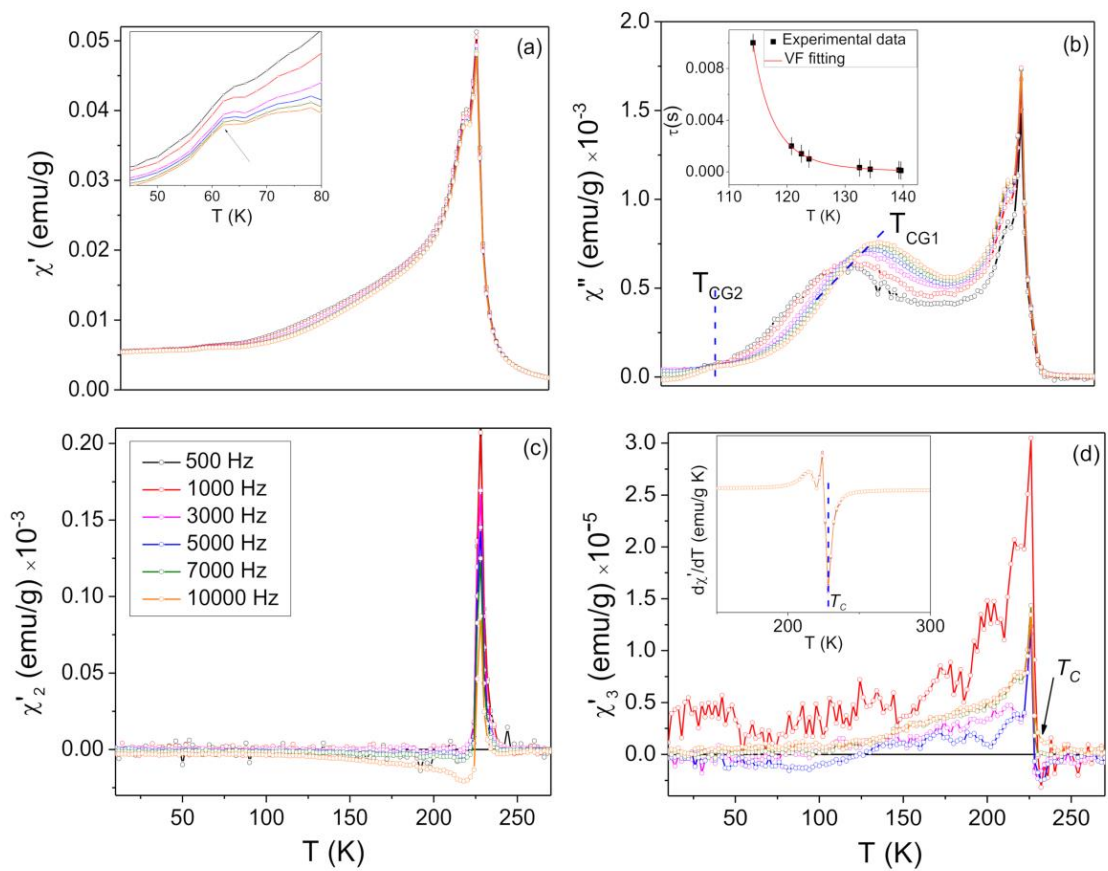

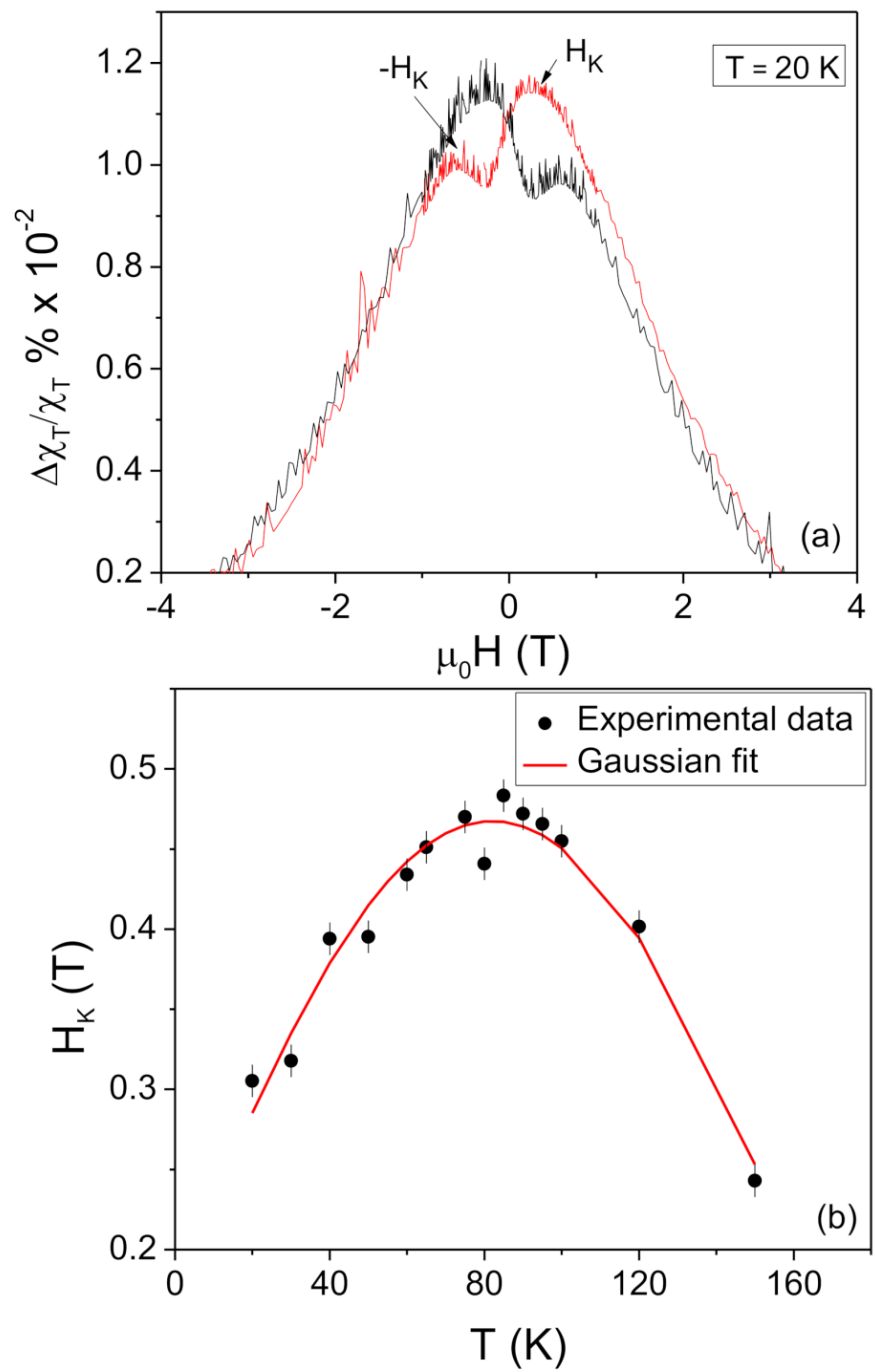

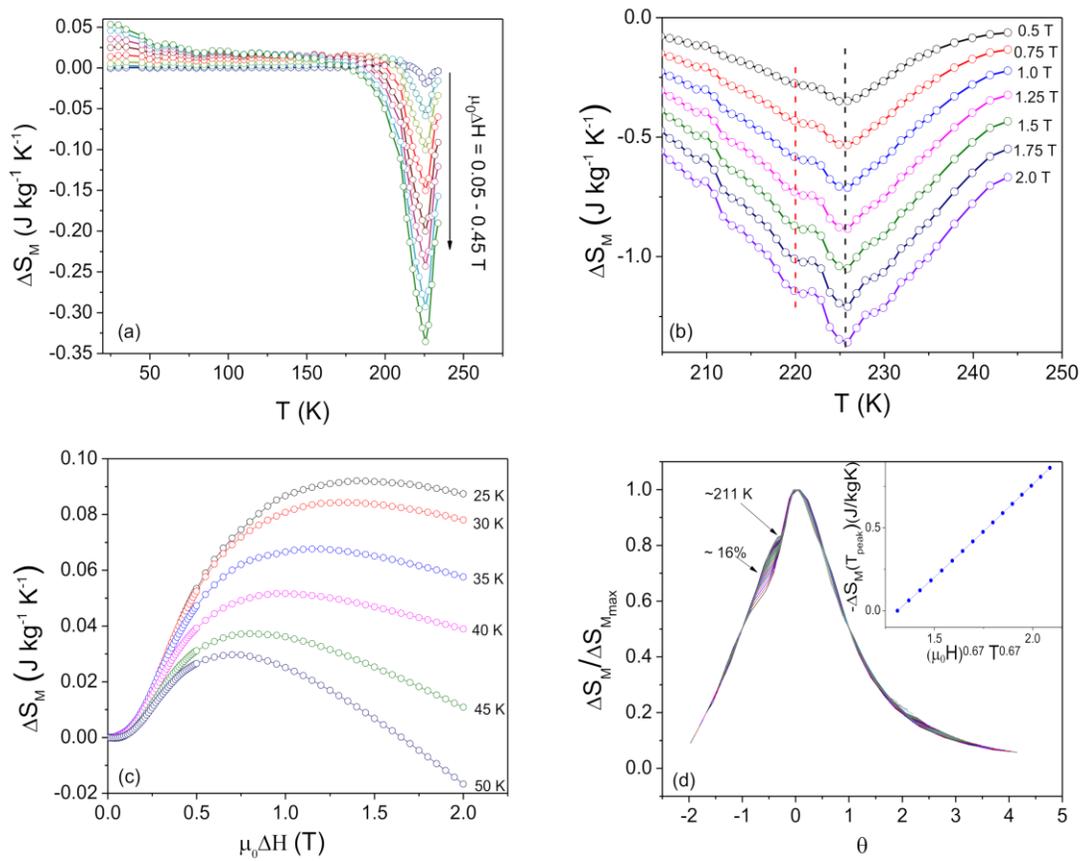

**FIG. 10**

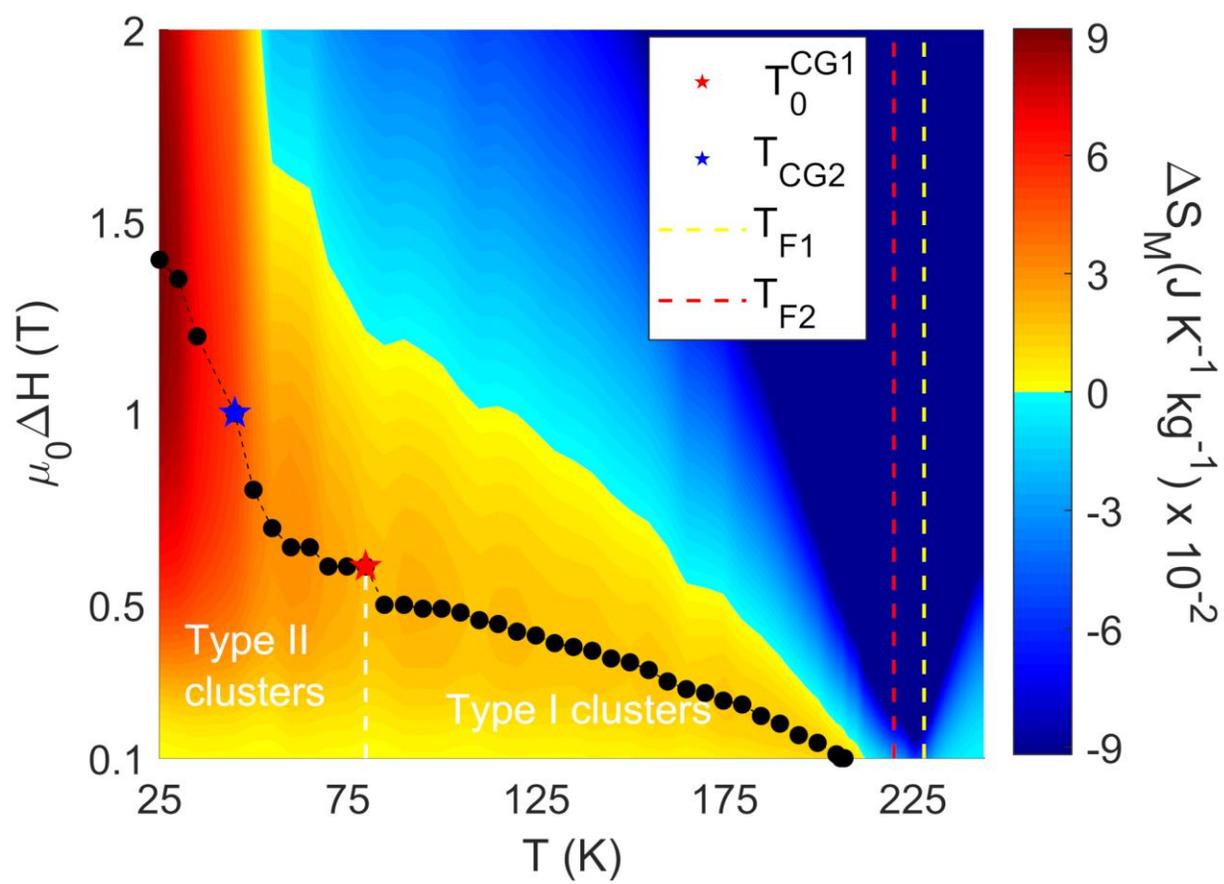